\documentclass[preprint,fleqn,showpacs,showkeys, nodayofweek]{revtex4}

\newcommand{\beqn}{\begin{eqnarray}}
\newcommand{\eeqn}{\end{eqnarray}}
\newcommand{\be}{\begin{equation}}
\newcommand{\ee}{\end{equation}}

\usepackage{graphicx}
\usepackage{amssymb}
\usepackage{amsmath}
\usepackage{color}
\usepackage{bm}
\usepackage{datetime}
\begin{document}
\title{Future cosmological evolution in $f(R)$ gravity using two equations of state parameters}

\author{Hyung Won Lee}
\email{hwlee@inje.ac.kr}

\author{Kyoung Yee Kim}
\email{kimky@inje.ac.kr}

\author{Yun Soo Myung}
\email{ysmyung@inje.ac.kr}

\affiliation{Institute of Basic Science and School of
Computer Aided Science, Inje University, Gimhae 621-749, Korea}

\begin{abstract}
We investigate  the issues of future oscillations around the phantom
divide for $f(R)$ gravity.  For this purpose, we introduce two types
of energy density and pressure arisen from the $f(R)$-higher order
curvature terms. One has the conventional energy density and
pressure even in the beginning of the Jordan frame, whose continuity
equation provides the native equation of state $w_{\rm DE}$. On the
other hand, the other   has the different forms of  energy density
and pressure which do not obviously satisfy the continuity equation.
This needs to introduce the effective equation of state $w_{\rm
eff}$ to describe the $f(R)$-fluid, in addition to the native
equation of state $\tilde{w}_{\rm DE}$. We confirm that future
oscillations around the phantom divide occur in $f(R)$ gravities by
introducing two types of equations of state. Finally, we point out
that the singularity appears ar $x=x_c$ because the stability
condition of $f(R)$ gravity violates.
\end{abstract}


 \shortdate
\maketitle

\section{Introduction}

Supernova type Ia(SUN Ia) observations has shown that our universe
is accelerating~\cite{SN}. Also cosmic microwave background
radiation~\cite{Wmap}, large scale structure~\cite{lss}, and weak
lensing~\cite{wl} have indicated that the universe has been
undergoing an accelerating phase since the recent past. The standard
model of ${\rm \Lambda}$CDM enables to explain these observational
results within observational error bound.  However, this model
suffers from the cosmological constant problem and thus, one needs
to find another model. Up to now, the $f(R)$-gravity as a modified
gravity model remains  a promising model  to explain the present
accelerating universe~\cite{cst,NO,sf,NOuh,FT}.
$f(R)$ gravities can
be considered as Einstein gravity (massless graviton) with an
additional scalar.   For example, it was shown that the
metric-$f(R)$ gravity is equivalent to the $\omega_{\rm BD}=0$
Brans-Dicke (BD) theory with the potential~\cite{FT}. Although the
equivalence principle test  in the solar system imposes a strong
constraint on $f(R)$ gravities, they may not be automatically ruled
out if the Chameleon mechanism is introduced to resolve it in the Einstein frame. It was
shown that the  equivalence principle test allows $f(R)$ gravity
models that are indistinguishable from the ${\rm \Lambda}$CDM model
in the background universe evolution~\cite{PS}.

In order to point out  the difference between  ${\rm \Lambda}$CDM
and $f(R)$ gravity, it is necessary to introduce the equation of
state parameter $w_{\rm DE}$.  Working  with the $f(R)$-gravity
action in the Jordan frame~\cite{Jordan}, one has to use  the
different energy density and pressure $(\tilde{\rho}_{\rm DE},
\tilde{p}_{\rm DE}) $ in compared to $(\rho_{\rm DE},p_{\rm DE})$ in
the Einstein-like frame~\cite{BGL,MSY}.  This corresponds to the
scalar-tensor (Brans-Dicke) theory in the Jordan frame~\cite{LKM}.
In the Einstein-like frame, one needs only the native equation of
state $w_{\rm DE}=p_{\rm DE}/\rho_{\rm DE}$ as in  the scalar-tensor
(quintessence model) theory, while one requires two equations of
states: ${\tilde w}_{\rm DE}={\tilde p}_{\rm DE}/{\tilde \rho}_{\rm
DE}$ and the effective equation of state $w_{\rm eff}$~\cite{ZP1}
because of non-minimal coupling of scalar to the gravity.  It  is
worth noting  that there is an essential difference between
Einstein-like and Einstein frames because the latter is recovered
from the conformal transformation in Jordan frame~\cite{FT,PHSS}.
For the holographic dark energy model, two of authors have clarified
that although there is a phantom phase when using the native
equation of state~\cite{WGA}, there is no phantom phase when using
the effective equation of state~\cite{KLM,KLMb}.

Recently, there were a few of important  works which explain the
oscillation around the future de Sitter solution with $w_{\rm
dS}=-1$ using $f(R)$-gravity~\cite{BGL} and its Brans-Dicke
theory~\cite{LKM}. Interestingly, the authors in~\cite{MSY} have
shown that the number of phantom divide crossings are infinite when
using the Ricci scalar perturbation, which is confirmed by
analytical condition and numerical way.

In this work, we focus on the issues of future oscillations around
the phantom divide for $f(R)$ gravity.  In order to confirm the
appearance of future oscillations around the phantom divide $w_{\rm
dS}=-1$, we introduce two types of  equation of states $w_{\rm DE}
(\tilde{w}_{\rm DE})$ and $w_{\rm eff}$  arisen from the
$f(R)$-fluid. We clarify the difference between two different sets
of  energy density and pressure by observing the ``negative and
effective" equations of state. Finally, we point out that the
singularity appears at $x=x_c$ because the stability condition of
$f(R)$ gravity violates when $F'=f''=0$ at the certain point
$x=x_c$.

\section{Future volution with $f(R)$-fluid in Einstein-like frame  } \label{eq_of_motion}
We start from the action of $f(R)$ gravity with matter as
\begin{equation}
I = \frac{1}{2\kappa^2}\int d^4 x \sqrt{-g} f(R) + I_{\rm m}(g_{\mu\nu}, \psi_{\rm m}),
\label{action}
\end{equation}
where $f(R)$ is  a function of Ricci scalar $R$ with $\kappa^2=8\pi
G$ and $I_{\rm m}$ is the action for matter which is assumed to be
minimally coupled to gravity. Here the action $I$ is initially
written in Jordan frame and $\psi_{\rm m}$ denotes matters. Taking
the variation of the action (\ref{action}) with respect to metric
$g_{\mu\nu}$, one obtains
\begin{equation}
FG_{\mu\nu} = \kappa^2 T_{\mu\nu}^{\rm (m)}
- \frac{1}{2}g_{\mu\nu} \left ( FR - f \right ) + \nabla_\mu \nabla_\nu F
- g_{\mu\nu} \nabla^2 F,
\label{variation}
\end{equation}
where $G_{\mu\nu}=R_{\mu\nu}-\frac{g_{\mu\nu}}{2}R$ is the Einstein
tensor and $F(R)=f'(R)$. Assuming the flat
Friedmann-Roberston-Walker (FRW) universe
\begin{equation}
ds^2_{\rm FRW} = -dt^2 +a^2(t) \Big( dr^2 + r^2d\Omega^2_2 \Big)
\label{metric}
\end{equation}
with $a(t)$ is the scale factor, we obtain the two Friedmann
equations from (\ref{variation}):
\begin{eqnarray}
3FH^2 &=& \kappa^2 \rho_{\rm M} + \frac{1}{2} \left ( FR - f \right ) -3 H {\dot F},\label{hubble}\\
-2F {\dot H} &=& \kappa^2 \left ( \rho_{\rm M} + p_{\rm M} \right ) + {\ddot F} - H {\dot F}, \label{second-hubble}
\end{eqnarray}
where $H = {\dot a}/a$ is the Hubble parameter, the overdot denotes
the  derivative with respect to the cosmic time  $t$, and $\rho_{\rm
M}$ and $p_{\rm M}$ are the energy density and pressure of all
perfect fluid-type matter, respectively. On the other hand,  the
scalar curvature $R$ defined by
\begin{equation}
R = 6 \left ( {\dot H} + 2 H^2 \right ) \label{curvature}
\end{equation}
plays an independent  role in the cosmological evolution because we
are working with  $f(R)$-fluid.  For our purpose,  we introduce the
new variable $x = \ln a$, then (\ref{hubble}) and (\ref{curvature})
take the forms
\begin{eqnarray}
H^2 &=& \left ( F-1 \right ) \left ( H \frac{dH}{dx} + H^2 \right )
  -\frac{1}{6}\left ( f - R \right ) - H^2 F' \frac{dR}{dx}
  + \frac{\kappa^2 \rho_m^0e^{-3x}}{3} + \frac{\kappa^2 \rho_m^0 e^{-4x}}{3} \chi,
\label{hubble1}\\
R &=& 6 \left ( H \frac{dH}{dx} + 2 H^2 \right ),
\label{curvature1}
\end{eqnarray}
where $\rho_m^0$ is the current density of cold dark matter (CDM)
and  $\chi=\rho_r^0/\rho_m^0 \simeq 3.1 \times 10^{-4}$ is the
current density ratio of radiation and dark matter. Regarding
(\ref{hubble1}) as the evolution equation, we rewrite it as a
compact form
\begin{equation}
H^2 = \frac{\kappa^2 }{3} \left ( \rho_{\rm DE} + \rho_{\rm m} + \rho_{\rm r} \right ),
\label{hubble2}
\end{equation}
where $\rho_{\rm DE}$, $\rho_{\rm m}$, and $\rho_{\rm r}$ represent
dark energy, dark matter and radiation, respectively. Comparing
(\ref{hubble1}) with  (\ref{hubble2}) leads to a definition of dark
energy density arisen from the $f(R)$-gravity~\cite{BGL,MSY}
\begin{equation}
 \rho_{\rm DE} = \frac{3}{\kappa^2 } \Bigg[\left ( F-1 \right ) \left ( H \frac{dH}{dx} + H^2 \right )
  -\frac{1}{6}\left ( f - R \right ) - H^2 F' \frac{dR}{dx}\Bigg].
\label{dark-energy}
\end{equation}
This dark energy density satisfies the conservation law as
\begin{equation}
{\dot \rho}_{\rm DE} + 3H \left ( 1 + w_{\rm DE} \right ) \rho_{\rm
DE} = 0 \label{continuity}
\end{equation}
with the native equation of state
 \begin{equation} w_{\rm
DE}=\frac{p_{\rm DE}}{\rho_{\rm DE}}. \end{equation} Importantly, we
observe that even starting with the $f(R)$ gravity action in the
Jordan frame, we have manipulated it  so that   the Einstein
equation (\ref{variation}) is rewritten effectively as
\begin{equation}
G_{\mu\nu} =G_{\mu\nu}(1-F)+ \kappa^2 T_{\mu\nu}^{\rm (m)} -
\frac{1}{2}g_{\mu\nu} \left ( FR - f \right ) + \nabla_\mu
\nabla_\nu F - g_{\mu\nu} \nabla^2 F, \label{variation1}
\end{equation}
 to derive (\ref{hubble1}) and  (\ref{hubble2}) for obtaining the standard dark energy $\rho_{\rm DE}$  and pressure
$p_{\rm DE}$. Hence we call the solution to (\ref{hubble2}) with
(\ref{dark-energy})  as {\it cosmological evolution with
$f(R)$-fluid in Einstein-like frame}.

In order to solve (\ref{hubble1}) and (\ref{curvature1})
simultaneously, we define the convenient variables call the reduced
Ricci scalar $r$ and the present matter-density parameter
$\Omega_{\rm m}^0$
\begin{equation}
r \equiv \frac{R}{6H^2}, \ \ \
\Omega_{\rm m}^0 \equiv \frac{\rho_{\rm m}^0}{\rho_{\rm crit}^0} =
\frac{\kappa^2 \rho_{\rm m}^0}{3H_0^2},
\end{equation}
and density parameters
\begin{eqnarray}
\Omega_{\rm m} = \frac{\rho_{\rm m}}{\rho_{\rm crit}},~~ \Omega_{\rm
r}= \frac{\rho_{\rm r}}{\rho_{\rm crit}},~~ \Omega_{\rm DE} = 1-
\Omega_{\rm m} - \Omega_{\rm r}.
\end{eqnarray}
Then two  equations (\ref{hubble1}) and (\ref{curvature1}) can be
written  four  equations in terms of new variables
\begin{eqnarray}
\frac{d\Omega_{\rm m}}{dx} &=& -(2r-1)\Omega_{\rm m}, \label{Om-eq}\\
\frac{d\Omega_{\rm r}}{dx} &=& -2r\Omega_{\rm r}, \label{Or-eq}\\
\frac{dr}{dx} &=& -2r(r-2) + \frac{F}{6H^2F'} \left \{
-1 + \frac{\Omega_{\rm m} + \Omega_{\rm r}}{F} +r - \frac{1}{6H^2}\frac{f}{F}
\right \} \label{r-eq}, \\
\frac{dH}{dx} &=& (r-2)H \label{H-eq} .
\end{eqnarray}
Considering
 \begin{equation}
 \frac{d\Omega_{\rm DE}}{dx} = -2\Omega_{\rm DE} (r-2) +
\frac{\kappa^2}{3H^2} \frac{d\rho_{\rm DE}}{dx}, \end{equation}
 one obtains the native equation of states as functions of $r$ and
 density parameters as
\begin{equation}
w_{\rm DE}(r,\Omega_{\rm m},\Omega_{\rm r}) = -1 - \frac{2r-4 + 3 \Omega_{\rm m} +
4 \Omega_{\rm r}}{3(1-\Omega_{\rm m}-\Omega_{\rm r})} . \label{bare-eos-einstein}
\end{equation}
At this stage, we wish to comment on the tilde-definition for
density parameters used in Ref.~\cite{BGL}
\begin{eqnarray}
{\tilde \Omega}_{\rm m} &=& \frac{\rho_{\rm m}}{\rho_{\rm crit}^0}
= \Omega_{\rm m} h^2 , \\
{\tilde \Omega}_{\rm r} &=& \frac{\rho_{\rm m}}{\rho_{\rm crit}^0}
= \Omega_{\rm r} h^2 , \\
{\tilde \Omega}_{\rm DE} &=& \frac{\rho_{\rm m}}{\rho_{\rm crit}^0}
= \Omega_{\rm DE} h^2,
\end{eqnarray}
which are clearly  different from our definition by $h^2$ factor, which is defined
as $h(x)=H(x)/H_0$.
Finally, we mention that the initial conditions for $\Omega_{\rm m}, \Omega_{\rm r}, r$
and $H$ are given by
\begin{eqnarray}
\Omega_{\rm m}(0)=\Omega_{\rm m}^0,~~ \Omega_{\rm r}(0) = \chi
\Omega_{\rm m}^0,~~ r(0) = 2 + h_1,~~ H(0)= 1,
\end{eqnarray}
where $h_1$ is the time derivative of $H$ at the present time $t=0$ with $a(0)=1$. Then, $x\in[-\infty,\infty]$ with $x=0$ at the present time. We note
that $h_1$ is related to deceleration parameter $q$ as
\begin{equation}
h_1 = - \left ( 1 + q \right )
\end{equation}
with
\begin{equation}
q \equiv - \frac{{\ddot a}{a}}{{\dot a}^2} = - \frac{{\ddot a}}{a} \frac{1}{H^2}
= -\left . \frac{1}{H}\frac{dH}{dx} \right |_0 - 1
= - h_1 - 1.
\end{equation}

Now we study the cosmological evolution by choosing four  specific
models of $f(R)$-gravity with the native equation of state $w_{\rm
DE}$ only.

\subsection{Cosmological constant}
\begin{figure}[t!]
   \centering
   \includegraphics[width=0.9\textwidth]{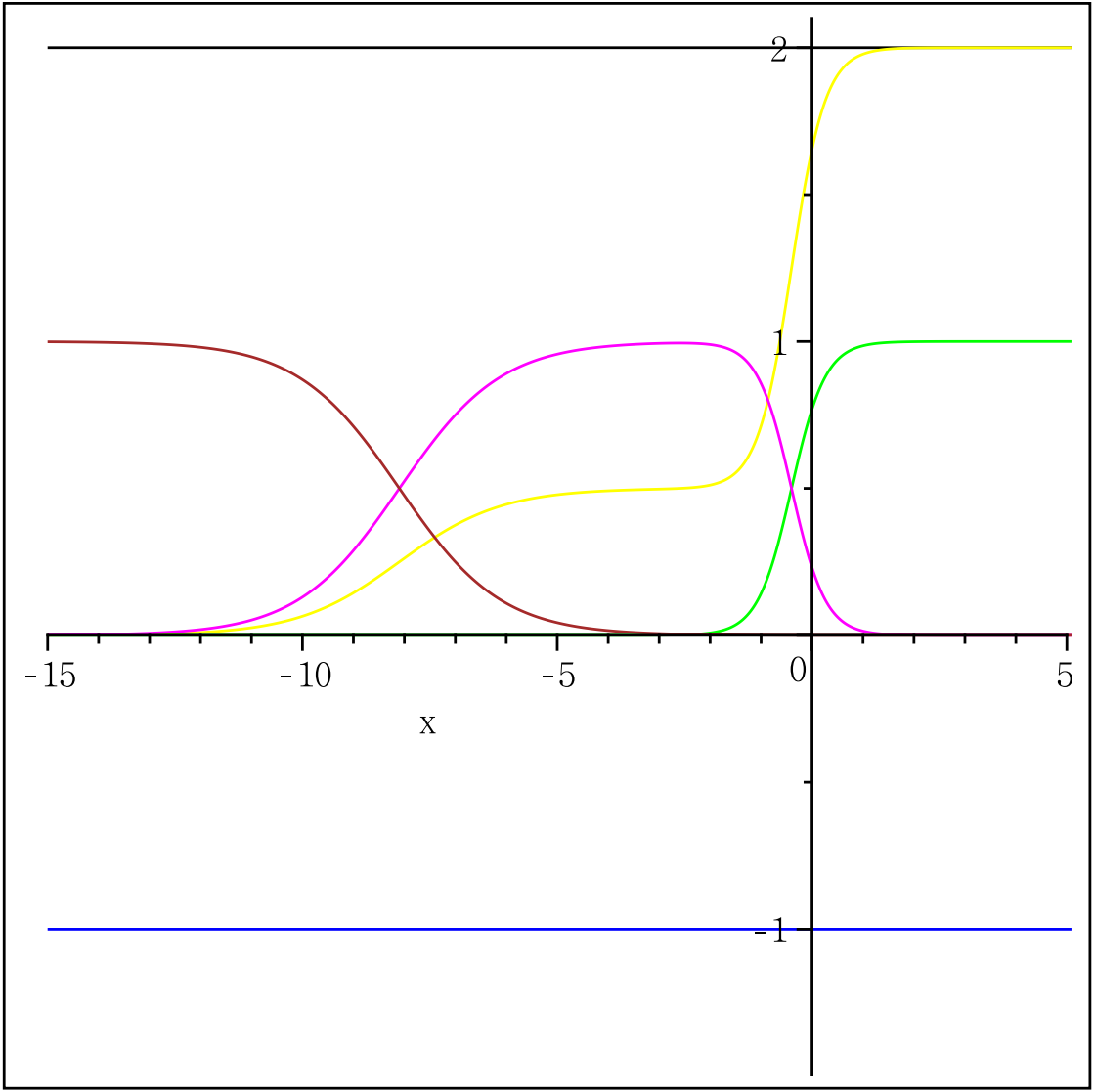}
   \caption{Time evolution of cosmological  parameters for cosmological constant case:
$w_{\rm DE}$(blue), $\Omega_{\rm DE}$(green), $\Omega_{\rm
m}$(magenta), $\Omega_{\rm r}$(brown), and $r$(yellow)  for
$\Omega_{\rm m}^0=0.23$ and $\chi = 3.04\times 10^{-4}, h_1 =
-(\frac{3}{2}+\frac{\chi}{2})\Omega_{\rm m}^0$. \label{fig1}}
\end{figure}

For cosmological constant case, the function $f(R)$ is simply  given
by
\begin{equation}
f(R) = R + \Lambda.
\end{equation}
In this case one cannot use (\ref{r-eq}) because of $F'(R)=f''(R)=0$, but
other equations are being  used to derive the solutions. Equation
(\ref{hubble1}) becomes
\begin{equation}
H^2 = - \frac{\Lambda}{6} + H_0^2 \Omega_{\rm m}^0 \left ( e^{-3x} + \chi e^{-4x} \right ).
\end{equation}
Differentiating this equation with respect to $x$, we get
\begin{equation}
2 H \frac{dH}{dx} = H_0^2 \Omega_{\rm m}^0 \left ( -3 e^{-3x} - 4 \chi e^{-4x} \right ).
\end{equation}
Plugging this into the definition of $r$, one gets
\begin{equation}
r = 2 - \frac{3}{2} \Omega_{\rm m} - 2 \Omega_{\rm r}.
\end{equation}
Hence, the relevant equation is just
\begin{eqnarray}
\frac{dr}{dx} = \frac{3}{2}(2r-1)\Omega_{\rm m} +4r \Omega_{\rm r}
\label{e-eq-constant}
\end{eqnarray}
with  $w_{\rm DE}= - 1$.
 Fig. \ref{fig1} depicts the  evolution for the cosmological
constant like the ${\rm \Lambda}$CDM without future oscillations
around the phantom divide $w_{\rm DE}=w_{\rm dS}=-1$ .
We point out that the reduced Ricci scalar  $r$ takes the value of
$r_{\rm dS}=2$ because of $R_{\rm dS}=12H^2$ for the de Sitter
spacetimes.
\subsection{Power-law gravity}
\begin{figure}[t!]
   \centering
   \includegraphics[width=0.9\textwidth]{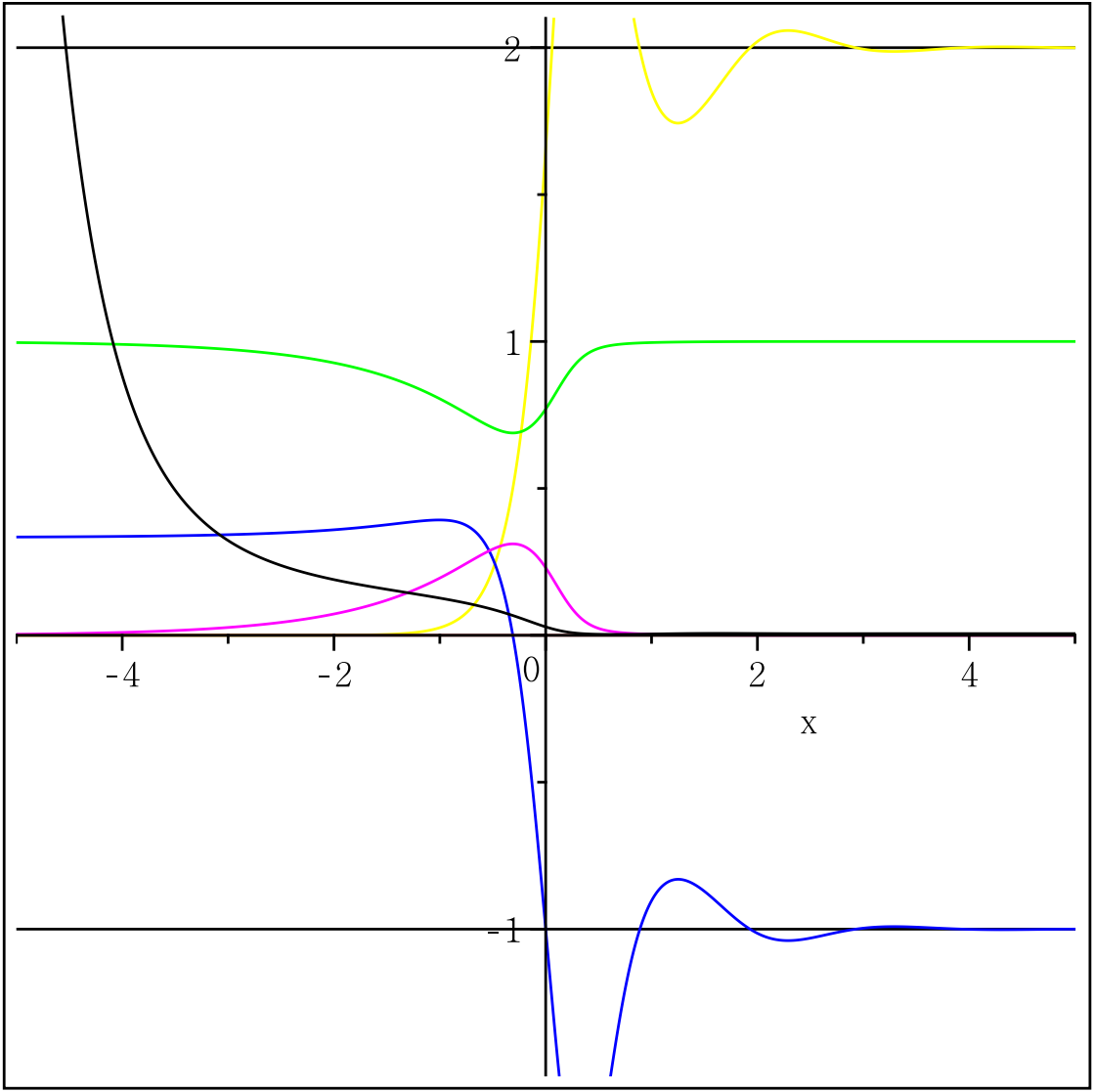}
   \caption{Time evolution of cosmological  parameters for power-law case:
$w_{\rm DE}$(blue), $\Omega_{\rm DE}$(green), $\Omega_{\rm
m}$(magenta), $\Omega_{\rm r}$(brown), $r$(yellow), and $F'$
(black) for ${f}_0=-3.6, \alpha=1/2, \Omega_{\rm m}^0=0.23,
\chi=3.04\times10^{-4}, h_1 =
-(\frac{3}{2}+\frac{\chi}{2})\Omega_{\rm m}^0$.\label{fig2}}
\end{figure}

\begin{figure}[t!]
   \centering
   \includegraphics[width=0.9\textwidth]{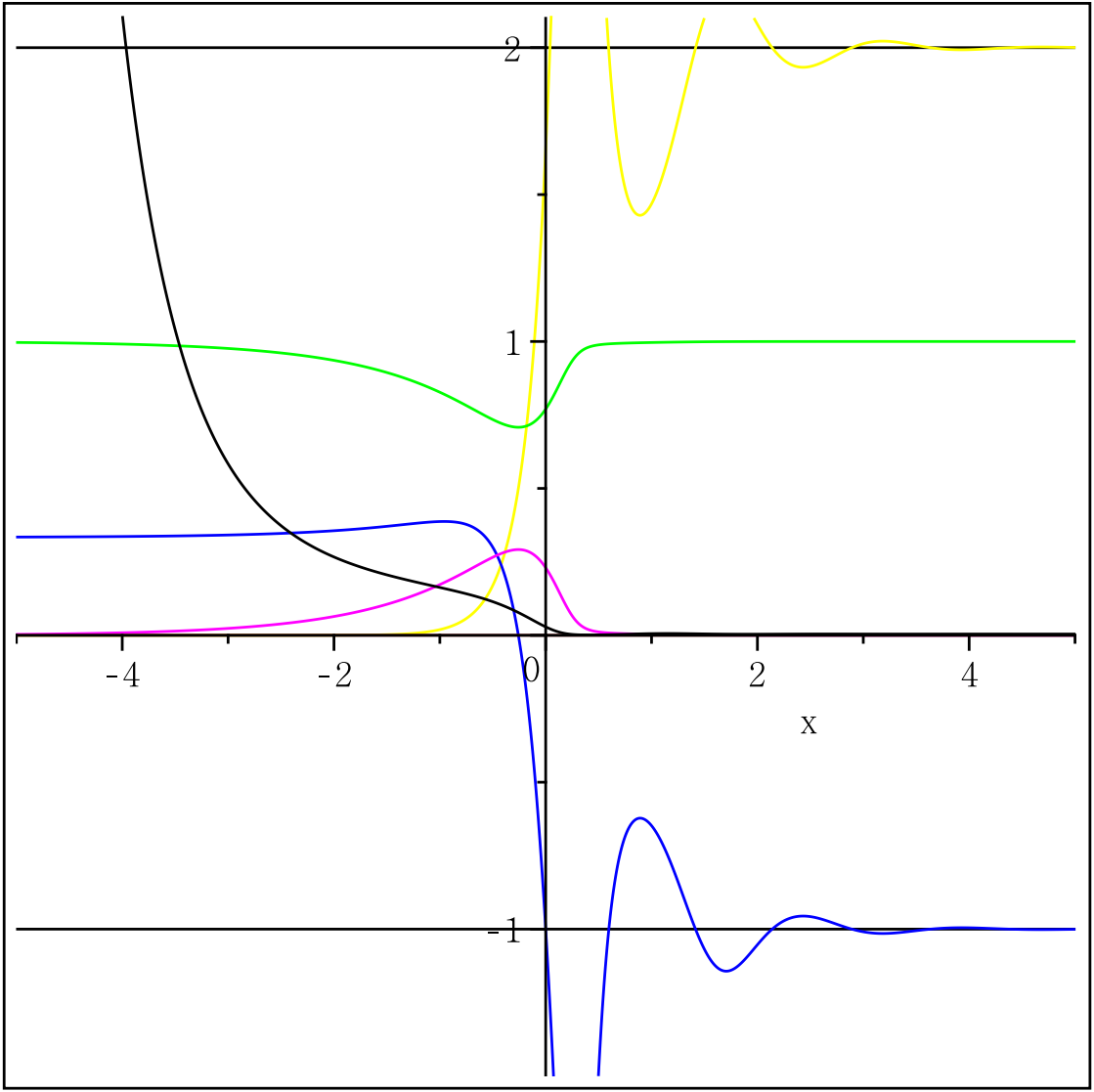}
   \caption{Time evolution of cosmological  parameters for power-law case:
$w_{\rm DE}$(blue), $\Omega_{\rm DE}$(green), $\Omega_{\rm
m}$(magenta), $\Omega_{\rm r}$(brown), $r$(yellow), and $F'$
(black) for ${f}_0=-6.5, \alpha=1/3, \Omega_{\rm m}^0=0.23,
\chi=3.04\times10^{-4}, h_1 =
-(\frac{3}{2}+\frac{\chi}{2})\Omega_{\rm m}^0$.\label{fig3}}
\end{figure}

When $f(R)$ takes the power-law form
\begin{equation} \label{p-law}
f(R) = R + f_0 R^{\alpha},
\end{equation}
Its derivatives with respect to $R$ are given by
\begin{eqnarray}
F(R) = 1 + \alpha f_0 R^{\alpha-1},~~ F'(R) = \alpha (\alpha-1) f_0
R^{\alpha-2}.
\end{eqnarray}
$w_{\rm DE}$ and $r$ in Figs. \ref{fig2} and \ref{fig3} represent
the future oscillations around the phantom divide for $\alpha=1/2$
and 1/3, respectively but they do not show  past  oscillations
around the phantom divide.

\subsection{Exponential gravity}
\begin{figure}[t!]
   \centering
   \includegraphics[width=0.9\textwidth]{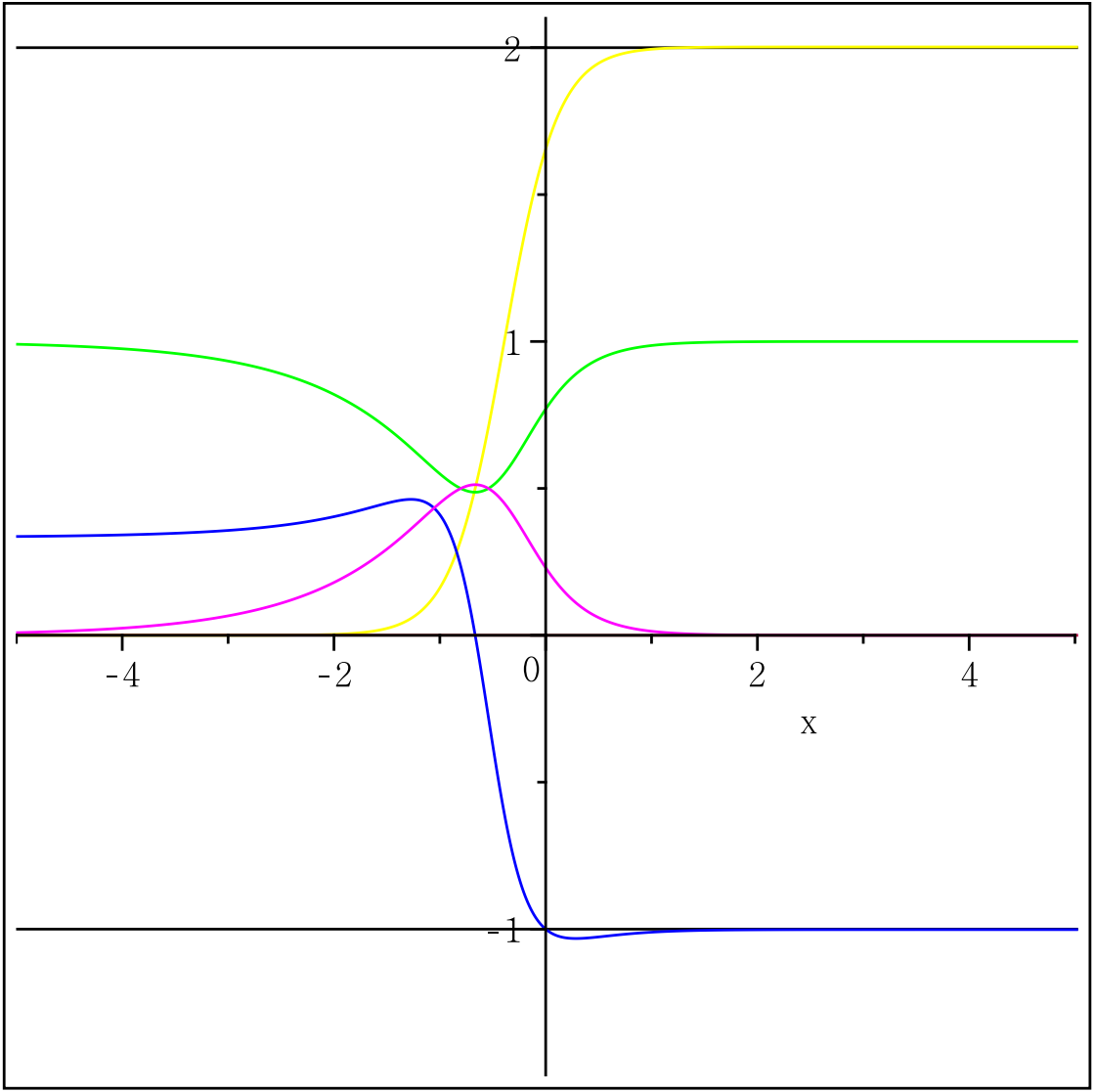}
   \caption{Time evolution of cosmological  parameters for an exponential case:
$w_{\rm DE}$(blue), $\Omega_{\rm DE}$(green), $\Omega_{\rm
m}$(magenta), $\Omega_{\rm r}$(brown), $r$(yellow),  and $F'$
(black) for $R_s=-0.05, \beta=1.1, \Omega_{\rm m}^0=0.23,
\chi=3.04\times10^{-4}, h_1 =
-(\frac{3}{2}+\frac{\chi}{2})\Omega_{\rm m}^0$.\label{fig4}}
\end{figure}

Now we wish to  apply the result of previous sub-sections to an
exponential gravity. Firstly, when  the function $f(R)$ is given by
\begin{equation} \label{ex-pot}
f(R) = R - \beta R_s \left ( 1 - e^{-R/R_s} \right ),
\end{equation}
its derivatives with respect to $R$ are given by
\begin{eqnarray}
F(R)=1 - \beta  e^{-R/R_s},~~ F'(R) = \frac{\beta}{R_s} e^{-R/R_s}.
\end{eqnarray}
Fig. \ref{fig4} indicates  no  the appearance of  future (past)
oscillations around the phantom divide for a given parameter
$R_s=-0.05$ and $\beta=1.1$ which is the same results found in
~\cite{BGLe}. We note that $F'$ does not appear in Fig. 4 because
its value is extremely large as $10^{87}$.

\subsection{Hu and Sawicki model} 
\begin{figure}[t!]
   \centering
   \includegraphics[width=0.9\textwidth]{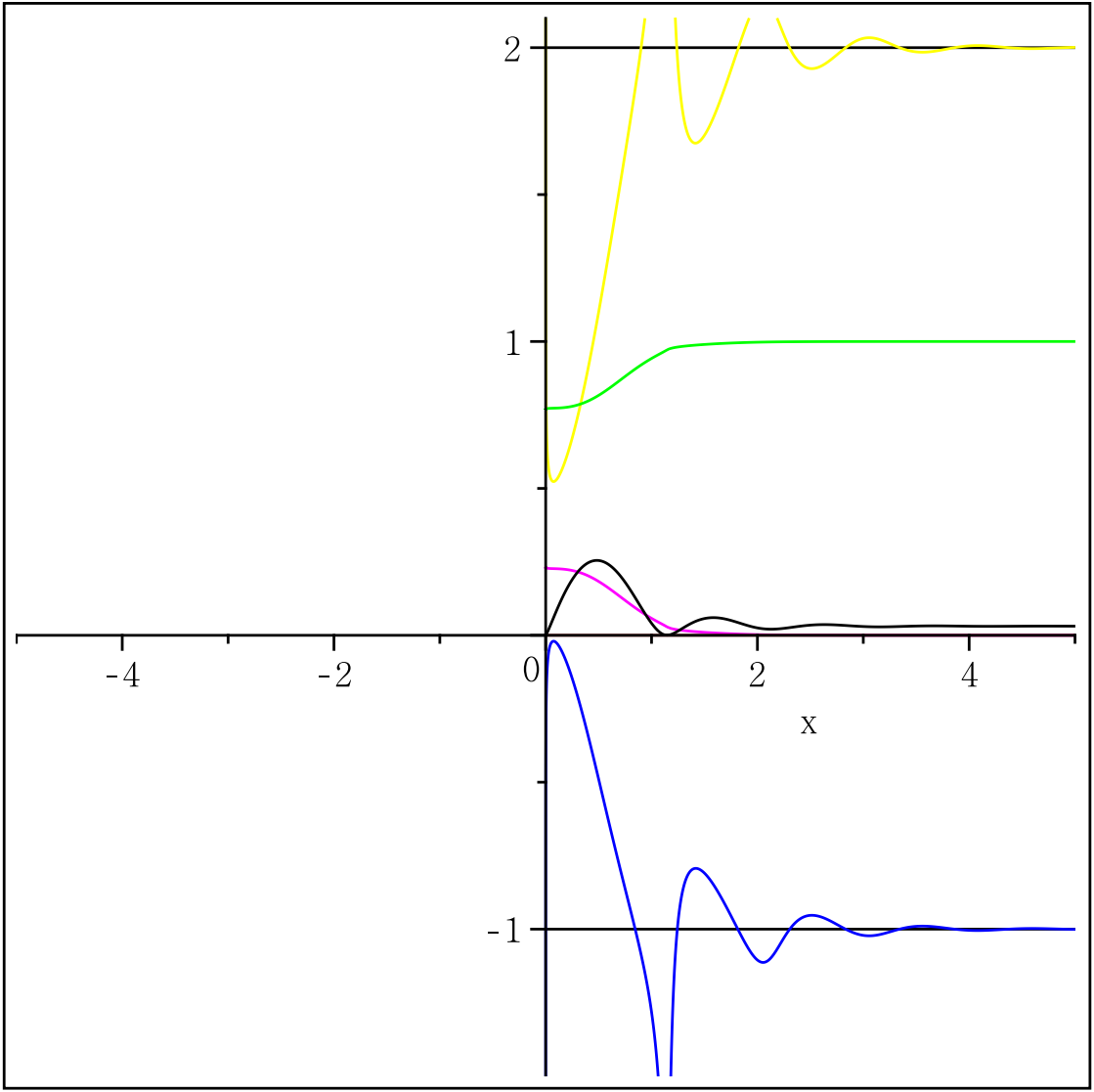}
   \caption{Time evolution of cosmological  parameters for the Hu and Sawicki model:
$w_{\rm DE}$(blue), $\Omega_{\rm DE}$(green), $\Omega_{\rm
m}$(magenta), $\Omega_{\rm r}$(brown), $r$(yellow),  and $F'$
(black) for $R_c=-1.0, \mu=-1.5, n = 2, \Omega_{\rm m}^0=0.23,
\chi=3.04\times10^{-4}, h_1 =
-(\frac{3}{2}+\frac{\chi}{2})\Omega_{\rm m}^0$.\label{fig5}}
\end{figure}

The Hu and Sawicki model takes the form
\begin{equation} \label{HSM}
 f(R) = R - \mu R_c \left [ 1 - \left (1 + \frac{R^2}{R_c^2}\right )^{-n}\right ].
 \end{equation}
Its derivatives with respect to $R$ are given by
\begin{eqnarray}
F(R) &=& 1 - 2 \mu n \frac{R}{R_c} \left ( 1+ \frac{R^2}{R_c^2}\right )^{-(n+1)}, \\
F'(R) &=& -\frac{2 \mu n }{R_c}
\left [
1- (2n+1) \frac{R^2}{R_c^2}
\right ]
\left ( 1+ \frac{R^2}{R_c^2}\right )^{-(n+2)} .
\end{eqnarray}
Fig. \ref{fig5} shows that  future oscillations around the phantom
divide $w_{\rm dS}=-1$ appears for $R_c=-1.0,~\mu=-1.5,$ and $n=2$,
but there is no evolution toward the past direction from $x_c=0$.
Similarly, there are future oscillations around $r_{\rm dS}=2$ for
the reduced Ricci scalar $r$.  We will explain why the evolution to
the past is not allowed in the Hu and Sawicki model in section IV.

\section{Cosmological evolution in Jordan frame}
In this section we derive evolution equations  in Jordan frame
without manipulation.  From equations  (\ref{hubble}) and
(\ref{second-hubble}), we have
\begin{eqnarray}
H^2 &=& \frac{1}{3F}\Bigg[\kappa^2 \rho_{\rm M} + \frac{1}{2} \left ( FR - f \right ) -3 H {\dot F}\Bigg],\label{hhjj}\\
- {\dot H} &=& \frac{1}{2F}\Bigg[\kappa^2 \left ( \rho_{\rm M} +
p_{\rm M} \right ) + {\ddot F} - H {\dot F}\Bigg], \label{ssjj}
\end{eqnarray}
Introducing  the reduced Ricci scalar $r$
\begin{equation}
\label{variables}
r = \frac{R}{6H^2},
\end{equation}
we can obtain an important  relation
\begin{eqnarray}
\frac{F'}{F}\frac{dR}{dx} &=& -1 +\Omega_{\rm m} +\Omega_{\rm r} +\frac{1}{6H^2}\left (R-\frac{f}{F}\right ), \label{drOverdx}
\end{eqnarray}
In this case, we read off  energy density and pressure of the dark
energy component from (\ref{hhjj}) and (\ref{ssjj})
as~\cite{Jordan}
\begin{eqnarray}
\tilde{\rho}_{\rm DE} &=& \frac{1}{\kappa^2 F} \left \{ \frac{FR-f}{2} -3 H {\dot F} \right \},
\label{density_de} \\
\tilde{p}_{\rm DE} &=& \frac{1}{\kappa^2 F} \left \{ -\frac{FR-f}{2} + 2 H {\dot F} + {\ddot F} \right \}.
\label{pressure_de}
\end{eqnarray}
Rewriting  the Einstein equation (\ref{variation}) as
\begin{equation}
G_{\mu\nu} = \kappa^2 \frac{T_{\mu\nu}^{\rm(m)}}{F} + \kappa^2 T_{\mu\nu}^{\rm DE},
\end{equation}
we obtain the non-conservation of continuity relation  by requiring
the Bianchi idensity
\begin{equation}
{\dot {\tilde\rho}}_{\rm DE} + 3 H \left ( \tilde{\rho}_{\rm DE} + \tilde{p}_{\rm DE}\right ) = \frac{\dot F}{F^2} \rho_{\rm M}.
\end{equation}
Because of the non-zero coupling term between $\rho_M$ and $\frac{\dot F}{F^2}$, we
must define an  ``effective" equation of state for $f(R)$-fluid
\begin{equation}
w_{\rm eff} = {\tilde w}_{\rm DE} -
\frac{-1+\Omega_{\rm m} + \Omega_{\rm r} +r -\frac{f}{6H^2F}}{3}
\Bigg[\frac{\Omega_{\rm m}+\Omega_{\rm r}}{1-\Omega_{\rm m}-\Omega_{\rm r}}\Bigg],
\end{equation}
with the native equation of state  $\tilde{w}_{\rm DE} = \tilde{p}_{\rm DE} /
\tilde{\rho}_{\rm DE}$.  This is similar to the Brans-Dicke
theory approach~\cite{KLM,LKM}. Defining density parameters newly as
\begin{equation}
\Omega_{\rm m} = \frac{\kappa^2}{3H^2}\frac{\rho_{\rm m}}{F}, \,\,
\Omega_{\rm r} = \frac{\kappa^2}{3H^2}\frac{\rho_{\rm r}}{F}, \,\,
\Omega_{\rm DE} = \frac{\kappa^2}{3H^2}\tilde{\rho}_{\rm DE},
\end{equation}
relevant quantities  are expressed by
\begin{eqnarray}
R &=& 6 H^2 r , \\
\tilde{\rho}_{\rm DE} &=& \frac{3H^2}{\kappa^2}
               \left \{ 1- \Omega_{\rm m} - \Omega_{\rm r} \right \}, \\
\tilde{p}_{\rm DE} &=& \frac{3H^2}{\kappa^2}
               \left \{-1 -\frac{1}{3}\Omega_{\rm r} -\frac{2}{3}(r-2) \right \} .
\end{eqnarray}
Insereting these into the definition of native equation of state, one gets
\begin{equation}
{\tilde w}_{\rm DE} = -1 -
\frac{2r-4+3\Omega_{\rm m} +4\Omega_{\rm r}}{3(1-\Omega_{\rm m} -\Omega_{\rm r})},
\label{bare-eos-jordan}
\end{equation}
which is the same as Eq. (\ref{bare-eos-einstein}) but their forms
of $\Omega_{\rm m}$ and $\Omega_{\rm m}$ are different  from those
in (\ref{bare-eos-einstein}).

Four equations  to be solved become
\begin{eqnarray}
\frac{d\Omega_{\rm m}}{dx} &=& -\left ( 2 r -2  + \Omega_{\rm m} + \Omega_{\rm r} + r -\frac{f}{6H^2F} \right )\Omega_{\rm m} , \label{omegam-eq-jordan}\\
\frac{d\Omega_{\rm r}}{dx} &=& -\left ( 2 r  -1 +\Omega_{\rm m} + \Omega_{\rm r} + r -\frac{f}{6H^2F} \right )\Omega_{\rm r} , \label{omegar-eq-jordan}\\
\frac{dr}{dx} &=& -2r(r-2) + \frac{F}{6H^2F'}
    \left ( -1 + \Omega_{\rm m} + \Omega_{\rm r} + r -\frac{f}{6H^2F} \right ), \label{r-eq-jordan}\\
\frac{dH}{dx} &=& (r-2) H \label{h-eq-jordan}.
\end{eqnarray}
Finally, we have to consider the initial conditions
\begin{equation}
H(0)=H_0 = 1, \,\, \Omega_{\rm r}(0)=\chi \Omega_{\rm m}^0, \,\,
r_0 = 2 + h_1, h_1 = -(1+q),
\end{equation}
with $q$ the deceleration parameter.

 Now we solve the above four
differential equations together with initial conditions by selecting
four interesting models.

\subsection{$f(R)=R+\Lambda$ case}
\begin{figure}[t!]
   \centering
   \includegraphics[width=0.9\textwidth]{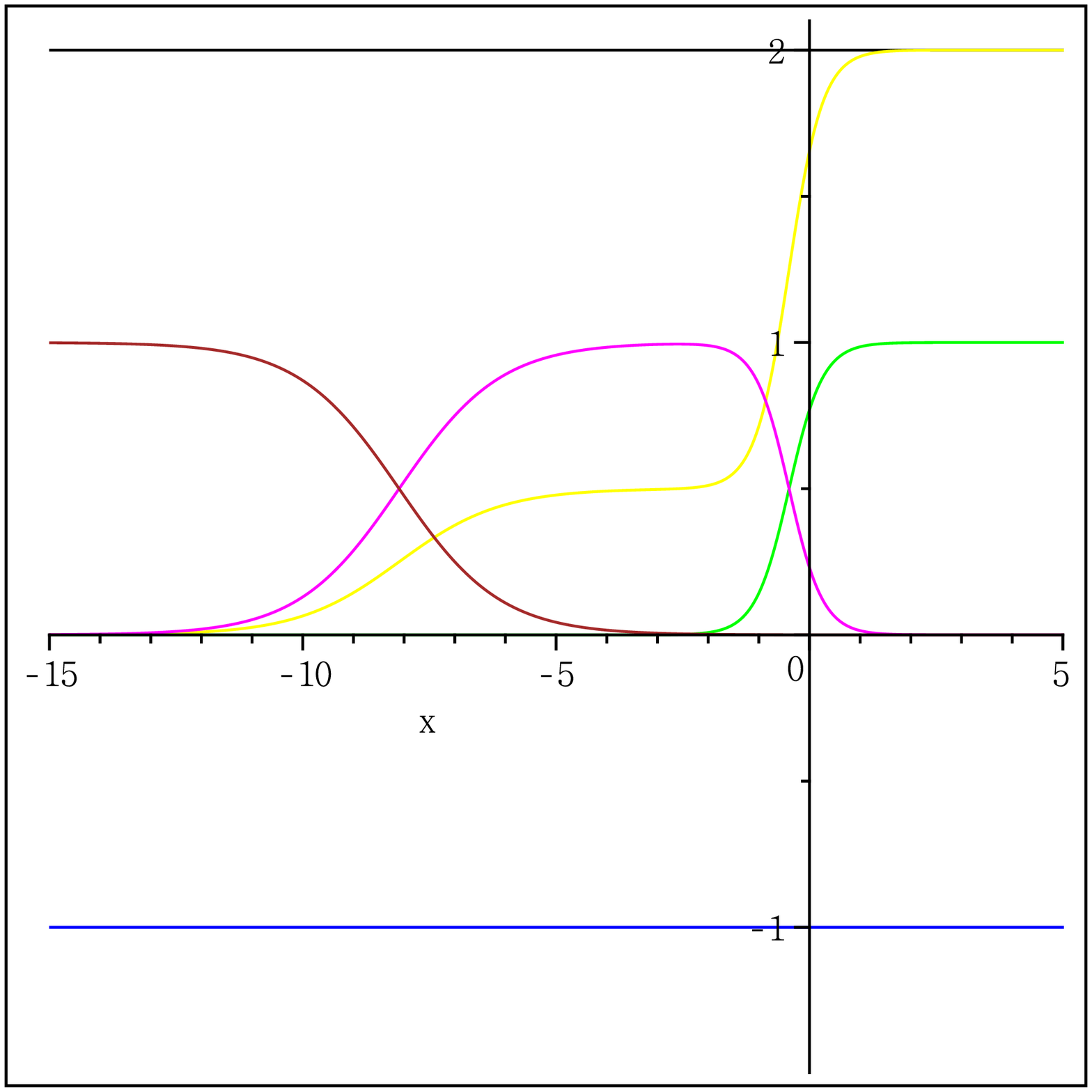}
   \caption{Time evolution of cosmological  parameters for cosmological constant case,
$w_{\rm DE}$(blue),
$\Omega_{\rm DE}$(green), $\Omega_{\rm m}$(magenta), $\Omega_{\rm r}$(brown), $r$(yellow), for
$\Omega_{\rm m}^0=0.23$ and $\chi = 3.04\times 10^{-4},
h_1 = -(\frac{3}{2}+\frac{\chi}{2})\Omega_{\rm m}^0$. \label{fig6}}
\end{figure}

In this case, we cannot use Eq. (\ref{r-eq-jordan}), which is valid only for
$F' \ne 0$.
In this case, $r$ is given by
\begin{equation}
r = 2 - \frac{3}{2}\Omega_{\rm m} -2 \Omega_{\rm r}.
\end{equation}
Hence the density and pressure of  $f(R)$-fluid are given
\begin{eqnarray}
\tilde{\rho}_{\rm DE} &=& \frac{3H^2}{\kappa^2} \left ( 1 - \Omega_{\rm m} - \Omega_{\rm r}\right ), \\
\tilde{p}_{\rm DE} &=& \frac{3H^2}{\kappa^2} \left ( -1 + \Omega_{\rm m} + \Omega_{\rm r}\right ).
\end{eqnarray}
Therefore, we have $\tilde{w}_{\rm DE} = -1= w_{\rm eff}$.
\begin{eqnarray}
\frac{d\Omega_{\rm m}}{dx} &=& -\left ( 2 r -1 \right )\Omega_{\rm m} , \label{omegam-eq-jordan-constant}\\
\frac{d\Omega_{\rm r}}{dx} &=& - 2 r  \Omega_{\rm r} , \label{omegar-eq-jordan-constant}\\
r &=& -\frac{3}{2}\Omega_{\rm m} - 2 \Omega_{\rm r}, \label{r-eq-jordan-constant}\\
\frac{dH}{dx} &=& (r-2) H \label{h-eq-jordan-constant}.
\end{eqnarray}
Note that these equations are exactly the same as the previous
section. Fig. \ref{fig6} shows the same result that  the  future
oscillations around the phantom divide  does not appear in the
Jordan frame.
We observe that the deceleration parameter $q$ is  fixed by the relation
\begin{equation}
1+q = - \left ( \frac{3}{2}  + \frac{\chi}{2} \right ) \Omega_{\rm m}^0.
\end{equation}
For $\Omega_{\rm DE}^0 = 0.77, \Omega_{\rm m}^0 = 0.23, \chi = 3.1\times 10^{-4}$, it
gives us $q = -0.6549643500$ and $ h_1 = -0.3451398400$.

\subsection{$f(R)=R+f_0 R^\alpha$ case }
\begin{figure}[t!]
   \centering
   \includegraphics[width=0.9\textwidth]{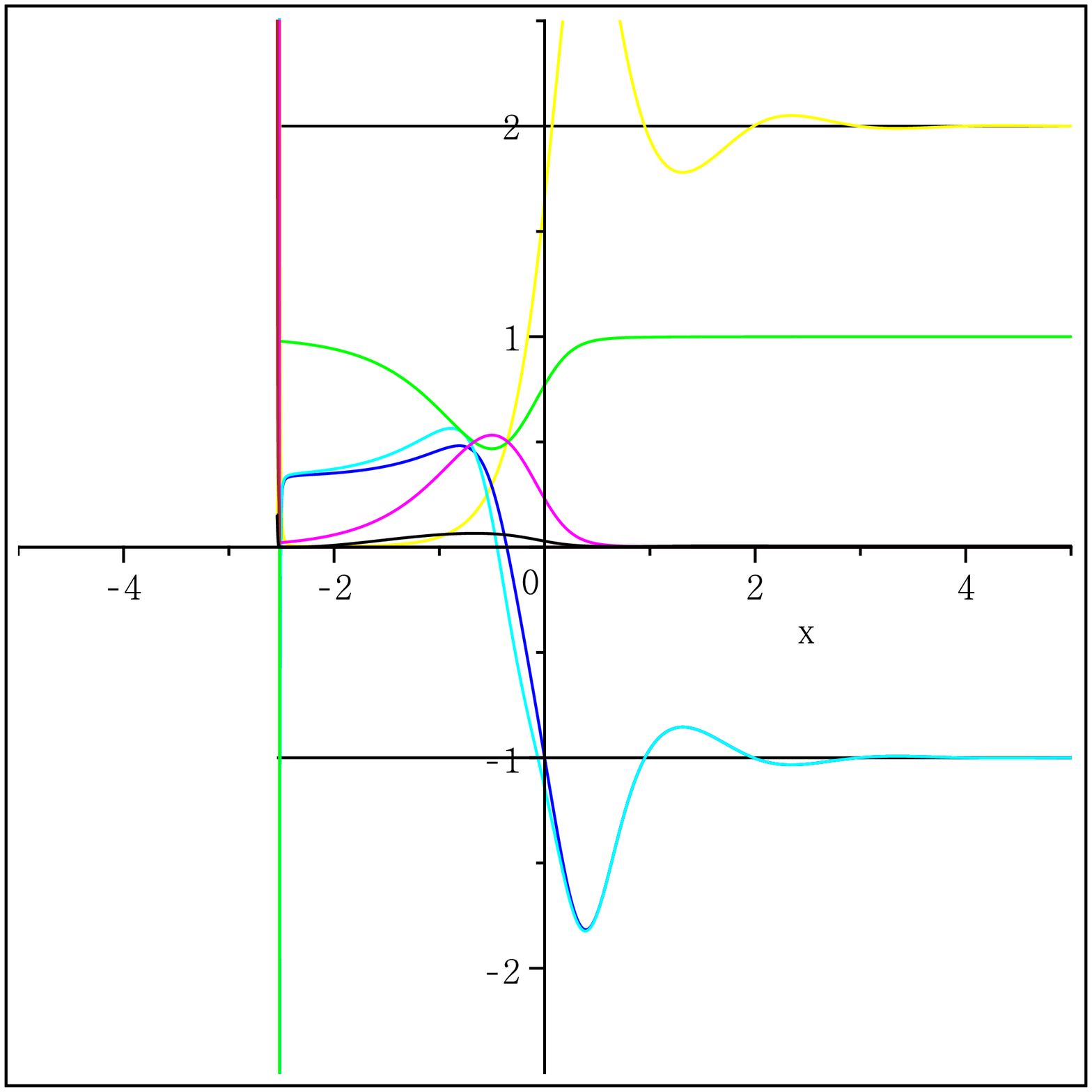}
   \caption{Time evolution of  cosmological  parameters for power-law case in Jordan frame:
$\tilde{w}_{\rm DE}$(blue), $w_{\rm eff}$(cyan), $\Omega_{\rm
DE}$(green), $\Omega_{\rm m}$(magenta), $\Omega_{\rm r}$(brown),
$r$(yellow), and $F'$ (black) for ${f}_0=-3.6, \alpha=1/2,
\Omega_{\rm m}^0=0.23, \chi=3.04\times10^{-4}, h_1 =
-(\frac{3}{2}+\frac{\chi}{2})\Omega_{\rm m}^0$.\label{fig7}}
\end{figure}

\begin{figure}[t!]
   \centering
   \includegraphics[width=0.9\textwidth]{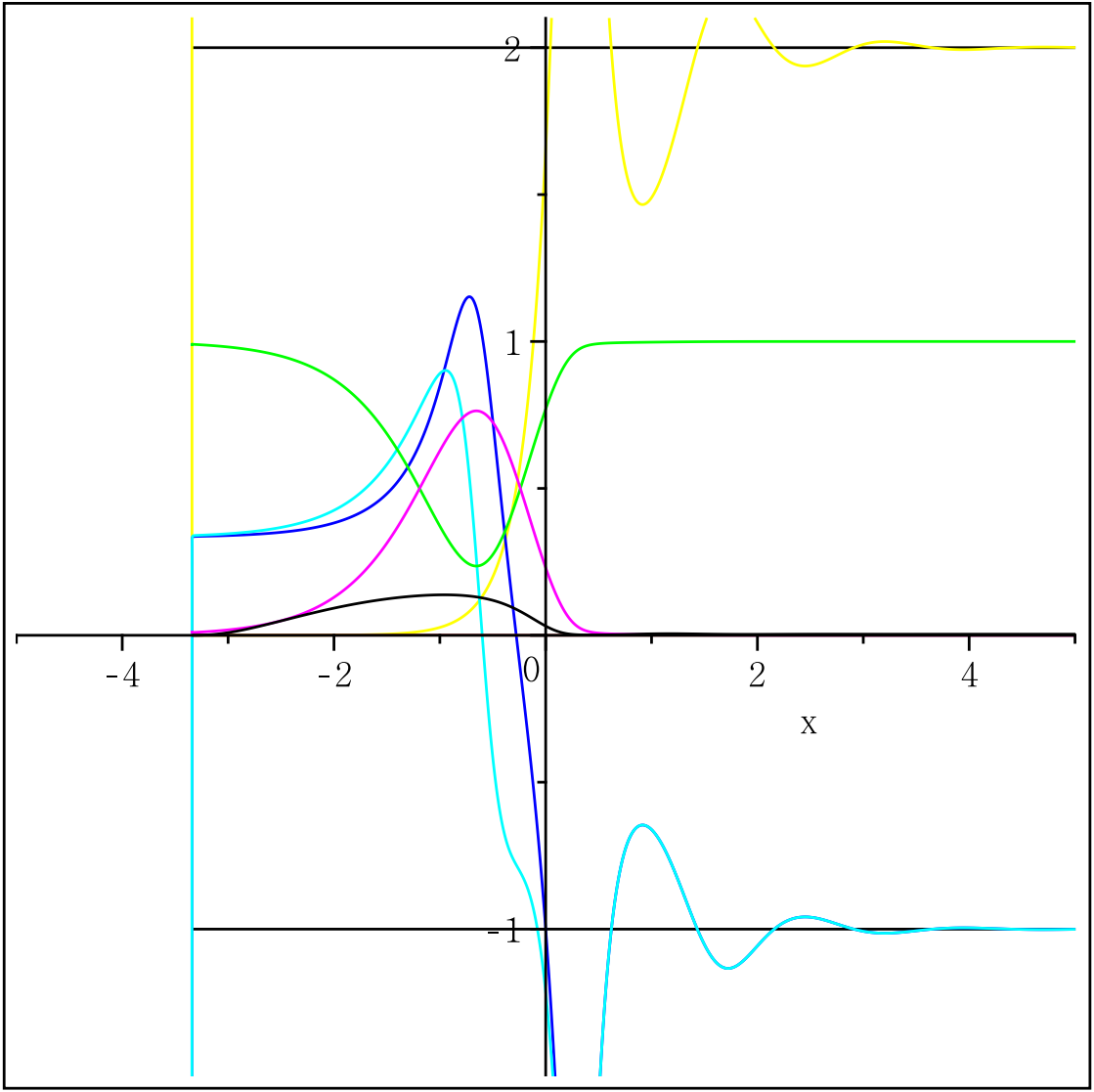}
   \caption{Time evolution of  cosmological  parameters for power-law case in Jordan frame:
$\tilde{w}_{\rm DE}$(blue), $w_{\rm eff}$(cyan), $\Omega_{\rm
DE}$(green), $\Omega_{\rm m}$(magenta), $\Omega_{\rm r}$(brown),
$r$(yellow), and $F'$ (black) for ${f}_0=-6.5, \alpha=1/3,
\Omega_{\rm m}^0=0.23, \chi=3.04\times10^{-4}, h_1 =
-(\frac{3}{2}+\frac{\chi}{2})\Omega_{\rm m}^0$.\label{fig8}}
\end{figure}

In this case, $F(R)$ is given
\begin{equation}
F(R) = 1 + \alpha f_0 R^{\alpha-1} .
\end{equation}
Its derivatives with respect to $R$ are given by
\begin{eqnarray}
F(R) &=& 1 + \alpha f_0 R^{\alpha-1}, \\
F'(R) &=& \alpha (\alpha-1) f_0 R^{\alpha-2} .
\end{eqnarray}
Figs. \ref{fig7} and \ref{fig8} show future oscillations around the
phantom divide using $\tilde{w}_{\rm DE}$  as in Figs. 2 and 3 but
the past evolution is terminated near $x_c \simeq -2.4$ for
$\alpha=1/2$ and $x_c\simeq 3.4$ for $\alpha=1/3$. Also, the reduced
Ricci scalar $r$ shows similar oscillating behaviors. This indicates
a violation of stability condition as will explain in section IV. We
confirm the appearance of future oscillations around the phantom
using the effective equation of state $w_{\rm eff}$.

\subsection{Exponential gravity}
\begin{figure}[t!]
   \centering
   \includegraphics[width=0.9\textwidth]{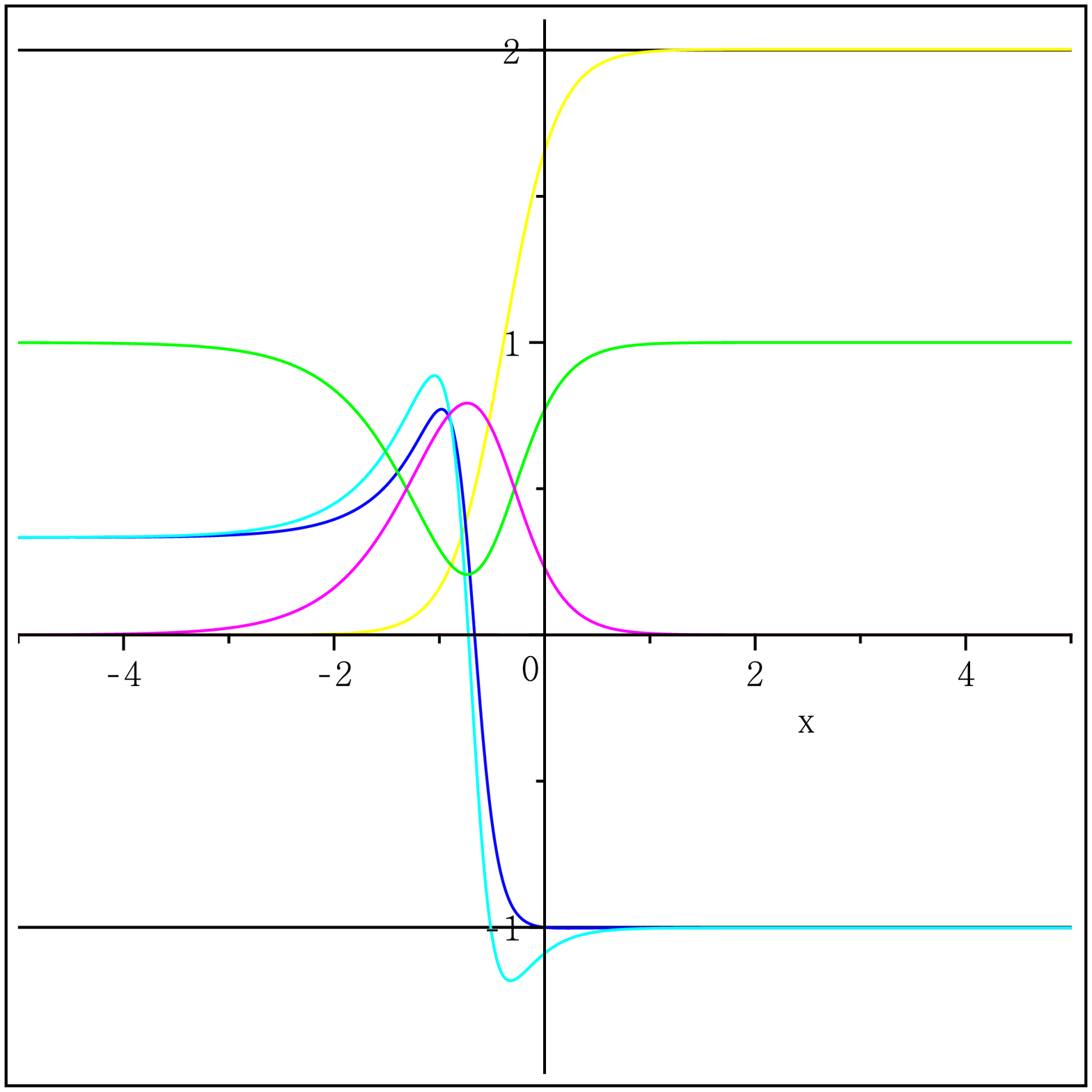}
   \caption{Time evolution of cosmological  parameters for exponential case in Jordan frame:
$\tilde{w}_{\rm DE}$(blue), $w_{\rm eff}$(cyan), $\Omega_{\rm
DE}$(green), $\Omega_{\rm m}$(magenta), $\Omega_{\rm r}$(brown),
$r$(yellow), and $F'$ (black) for $R_s=-0.05, \beta=1.1,
\Omega_{\rm m}^0=0.23, \chi=3.04\times10^{-4}, h_1 =
-(\frac{3}{2}+\frac{\chi}{2})\Omega_{\rm m}^0$.\label{fig9}}
\end{figure}

For the exponential gravity, the function $f(R)$ is given
\begin{equation}
f(R) = R - \beta R_s \left ( 1 - e^{-R/R_s} \right ).
\end{equation}
Its derivatives with respect to $R$ are given by
\begin{eqnarray}
F(R) &=& 1 - \beta  e^{-R/R_s}, \\
F'(R) &=& \frac{\beta}{R_s} e^{-R/R_s} .
\end{eqnarray}
Figs. \ref{fig9} depicts that future oscillations around
the phantom divide does not appear when using $\tilde{w}_{\rm DE}$ and $w_{\rm eff}$.
There is no essential  difference between Einstein-like (Fig. 4) and Jordan frames (Fig. 9).
 We would like to mention that $F'$ does not appear in Fig. 4 because
its value is extremely large as $10^{87}$.

\subsection{Hu and Sawicki case $f(R) = R - \mu R_c \left [ 1 - \left (1 + \frac{R^2}{R_c^2}\right )^{-n}\right ]$}

\begin{figure}[t!]
   \centering
   \includegraphics[width=0.9\textwidth]{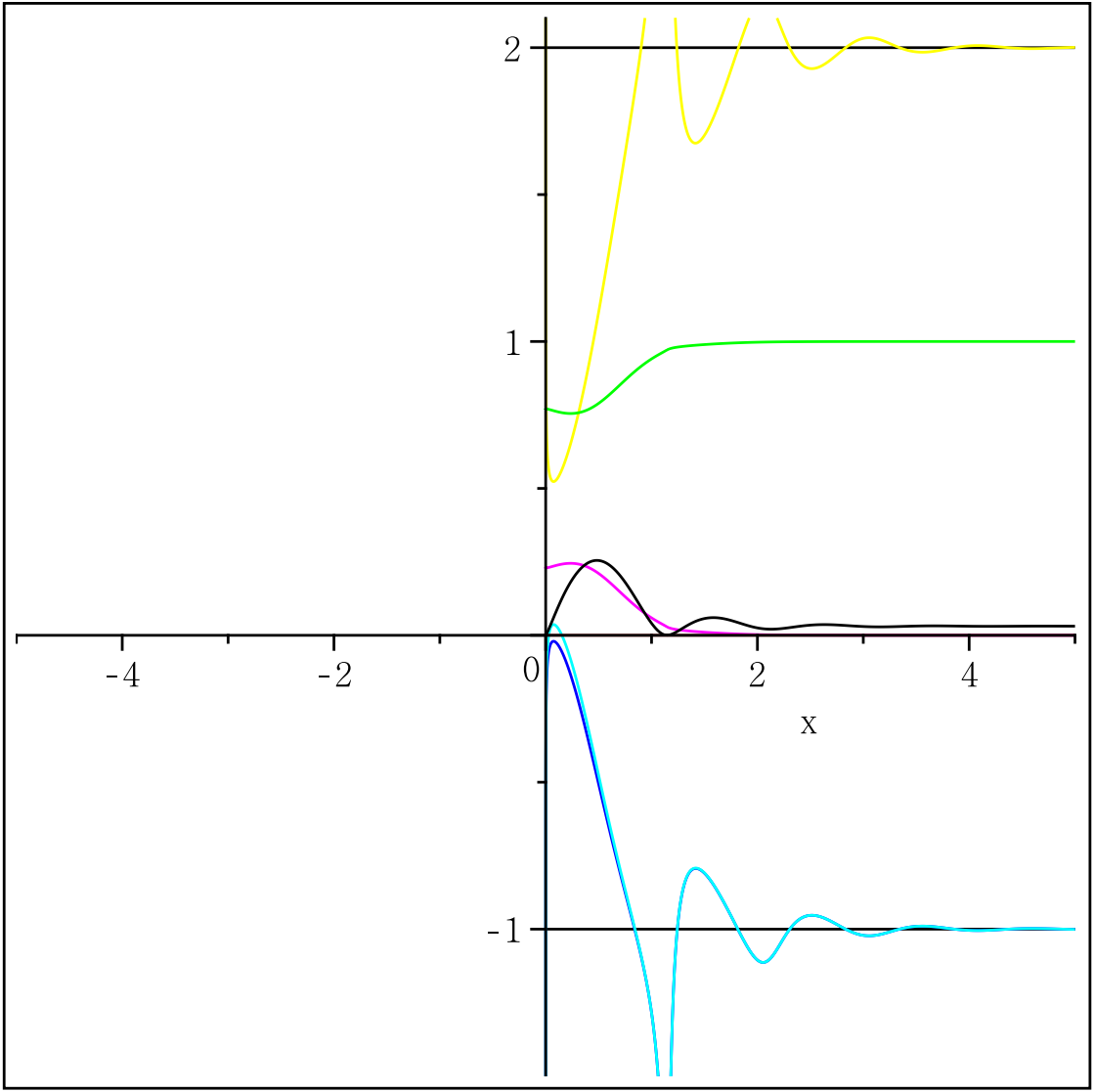}
   \caption{Time evolution of cosmological  parameters for Hu and Sawicki case in Jordan frame:
$\tilde{w}_{\rm DE}$(blue), $w_{\rm eff}$(cyan), $\Omega_{\rm
DE}$(green), $\Omega_{\rm m}$(magenta), $\Omega_{\rm r}$(brown),
$r$(yellow), and $F'$ (black) for $R_c=-1.0, \mu=-1.5, n=2,
\Omega_{\rm m}^0=0.23, \chi=3.04\times10^{-4}, h_1 =
-(\frac{3}{2}+\frac{\chi}{2})\Omega_{\rm m}^0$.\label{fig10}}
\end{figure}

Its derivatives with respect to $R$ are given by
\begin{eqnarray}
F(R) &=& 1 - 2 \mu n \frac{R}{R_c} \left ( 1+ \frac{R^2}{R_c^2}\right )^{-(n+1)}, \\
F'(R) &=& -\frac{2 \mu n }{R_c}
\left [
1- (2n+1) \frac{R^2}{R_c^2}
\right ]
\left ( 1+ \frac{R^2}{R_c^2}\right )^{-(n+2)} .
\end{eqnarray}
Fig. \ref{fig10} shows that  future oscillations around the phantom
divide $w_{\rm dS}=-1$ appears for $R_c=-1.0,~\mu=-1.5,$ and $n=2$
using two equations of state $\tilde{w}_{\rm DE}$ and $w_{\rm eff}$
but there is no evolution toward the past direction from $x_c=0$.
Also, there are future oscillations around $r_{\rm dS}=2$ for the
reduced Ricci scalar $r$. This will be explained in the next
section.  This confirms the results (Fig. 5) in the Einstein-like
frame.

\section{Singularity in cosmological evolutions}

In this section, we wish to explain the singularities encountered in
the cosmological evolution of $f(R)$-fluid~\cite{Frolov}. First of
all,  we mention that $f(R)$-gravity should satisfy the following
bounds:~\cite{MSY}
\begin{equation} \label{stcon}
f'(R)= F(R)>0,~~f''(R)=F'(R)>0.
\end{equation}
These are necessary to guarantee that the Newtonian gravity
solutions are stable  and  that the matter-dominated stage remains
an attractor with respect to an open set of neighboring cosmological
solutions in $f(R)$-gravity. In the perturbation theory, the former
is necessary to show that the gravity is attractive and and the
graviton is not a ghost, whereas the latter needs to ensure that the
scalaron of a massive curvature scalar does not have a tachyon. We
show how the singularities appear from the $f(R)$-fluid. From the
observation of two equations (\ref{r-eq}) and (\ref{r-eq-jordan})
which are equivalent to two first Friedmann equations, the second
term involves $F'(R)$ as the denominator. Hence, if $F'(R)=0$ at a
certain point of $x=x_c$, it gives rise to an singularity at which
some cosmological parameters blow up. This results from the
violation of stability condition $f(R)$ gravity: the second of
(\ref{stcon}).

In order to show the presence of singularities explicitly, we use
the graph of $F'(x)$ as a function of $x$.  From Fig. 5, we find
that the singularity appears at $x_c=0$ and thus, the backward
evolution is not allowed in Einstein-like frame. In the Jordan
frame, the power-law gravity shows the singularities at $x_c \simeq
-2.4$ for $\alpha=1/2$ (see Fig. 7) and $x_c \simeq -3.4$ for
$\alpha=1/3$ (see Fig. 8), while from Fig. 10,  we find that the
singularity appears at $x_c=0$ and thus, the backward evolution is
not allowed in Jordan frame. This completes the presence of
singularities in the cosmological evolution of $f(R)$-fluid.

\section{Discussions}
We have investigated  the issues of future oscillations around the
phantom divide for $f(R)$ gravity by  introducing two types of
energy density and pressure arisen from the $f(R)$-fluid. One has
the conventional energy density $\rho_{\rm DE}$ and pressure $p_{\rm
DE}$  even in the beginning of the Jordan frame, whose continuity
equation provides the native equation of state  $w_{\rm DE}=p_{\rm
DE} /  \rho_{\rm DE}$. Hence, we call this frame as the
Einstein-like frame.

 On the other hand, the other has the different
forms of energy density $\tilde{\rho}_{\rm DE}$ and  pressure
$\tilde{p}_{\rm DE}$ which do not obviously satisfy the continuity
equation. This needs to introduce the effective equation of state
$w_{\rm eff}$ to describe the $f(R)$-fluid precisely, in addition to
the native equation of state $\tilde{w}_{\rm DE}=\tilde{p}_{\rm
DE}/\tilde{\rho}_{\rm DE}$. We confirm that future oscillations
around the phantom divide always occur in $f(R)$-gravities by
introducing two types of $f(R)$-gravity: one is the power-law
potential (\ref{p-law}) with the exponent $\alpha=1/2$ and 1/3 and
the other is the Hu and Sawiciki model (\ref{HSM}). In the Jordan
frame,  the former did not show  past  oscillations around the
phantom divide and its evolution was terminated around
$x_c\simeq-2.4$ for $\alpha=1/2$. On the the other hand, the latter
did not provide any past evolution in both Einstein-like and Jordan
frames. Similarly, we confirm that   there are future oscillations
around $r_{\rm dS}=2$ for the reduced Ricci scalar $r$~\cite{MSY}.
As was expected, the cosmological constant model has  no
frame-dependence and we could not find any  future oscillations
around the phantom divide around $w_{\rm DE}=-1$ for the exponential
gravity in (\ref{ex-pot}).

For whole evolution from the past to future when imposing initial
conditions at the present time, the cosmological evolution is
allowed in the Einstein-like frame  better than in the Jordan frame.
This means that the cosmological evolution of $f(R)$-fluid
determined from its form of energy density and pressure, depending
on the given frame.  Also, it was proven that the termination
(singularity) appeared in cosmological evolution is closely related
to the form of $f(R)$-fluid for given frame. This has arisen from
$F'=0$ in the first Friedmann  equations (\ref{r-eq}) and
(\ref{r-eq-jordan}). As a result, it is so  because of the violation
of the stability condition (non-tachyon) of $f(R)$ gravity.

Consequently, we have successfully performed (whole) cosmological
evolution of $f(R)$ gravities by choosing two different state
variables of energy density and pressure, and pointed out why the
singularity appeared in the backward evolution when the initial
condition was chosen as the present time.

\begin{acknowledgments}
This work was supported by the National Research Foundation of
Korea(NRF) grant funded by the Korea government(MEST)
(No.2010-0028080).
\end{acknowledgments}


\begin{thebibliography}{3}
\bibitem{SN}S.~Perlmutter {\it et al.}  [Supernova Cosmology Project Collaboration],
  Astrophys.\ J.\  {\bf 517} (1999) 565
  [arXiv:astro-ph/9812133];
A.~G.~Riess {\it et al.}  [Supernova Search Team Collaboration],
  Astron.\ J.\  {\bf 116} (1998) 1009
  [arXiv:astro-ph/9805201].

\bibitem{Wmap}
D.~N.~Spergel {\it et al.}  [WMAP Collaboration],
  Astrophys.\ J.\ Suppl.\  {\bf 170} (2007) 377
  [arXiv:astro-ph/0603449].

\bibitem{lss}
  M.~Tegmark {\it et al.}  [SDSS Collaboration],
  Phys.\ Rev.\  D {\bf 69} (2004) 103501
  [arXiv:astro-ph/0310723];
D.~J.~Eisenstein {\it et al.}  [SDSS Collaboration],
  Astrophys.\ J.\  {\bf 633} (2005) 560
  [arXiv:astro-ph/0501171].

\bibitem{wl}
  B.~Jain and A.~Taylor,
  Phys.\ Rev.\ Lett.\  {\bf 91} (2003) 141302
  [arXiv:astro-ph/0306046].


\bibitem{NO}
  S.~Nojiri and S.~D.~Odintsov,
  eConf {\bf C0602061} (2006) 06
  [Int.\ J.\ Geom.\ Meth.\ Mod.\ Phys.\  {\bf 4}, 115 (2007)]
  [arXiv:hep-th/0601213].

\bibitem{cst}
 E.~J.~Copeland, M.~Sami and S.~Tsujikawa,
  Int.\ J.\ Mod.\ Phys.\  D {\bf 15} (2006) 1753
  [arXiv:hep-th/0603057].

\bibitem{sf}
  T.~P.~Sotiriou and V.~Faraoni,
  Rev.\ Mod.\ Phys.\  {\bf 82}, 451 (2010)
  [arXiv:0805.1726 [gr-qc]].


\bibitem{NOuh}
  S.~Nojiri and S.~D.~Odintsov,
  arXiv:1011.0544 [gr-qc].


\bibitem{FT}
  A.~De Felice and S.~Tsujikawa,
  Living Rev.\ Rel.\  {\bf 13}, 3 (2010)
  [arXiv:1002.4928 [gr-qc]].


\bibitem{PS}
L.~Pogosian and A.~Silvestri,
  Phys.\ Rev.\  D {\bf 77} (2008) 023503
  [Erratum-ibid.\  D {\bf 81}, 049901 (2010)]
  [arXiv:0709.0296 [astro-ph]].

\bibitem{Jordan}
 S.~Capozziello, M.~De Laurentis and V.~Faraoni,
  arXiv:0909.4672 [gr-qc];
  K.~Nozari and T.~Azizi,
  Phys.\ Lett.\  B {\bf 680} (2009) 205
  [arXiv:0909.0351 [gr-qc]].



\bibitem{BGL}
  K.~Bamba, C.~Q.~Geng and C.~C.~Lee,
  JCAP {\bf 1011}, 001 (2010)
  [arXiv:1007.0482 [astro-ph.CO]].

\bibitem{MSY}
  H.~Motohashi, A.~A.~Starobinsky and J.~Yokoyama,
  JCAP {\bf 1106} (2011) 006
  [arXiv:1101.0744 [astro-ph.CO]].


\bibitem{LKM}
  H.~W.~Lee, K.~Y.~Kim and Y.~S.~Myung,
  Eur.\ Phys.\ J.\  C {\bf 71}, 1585 (2011)
  [arXiv:1010.5556 [hep-th]].

  \bibitem{ZP1}
  W.~Zimdahl and D.~Pavon,
  Phys.\ Lett.\  B {\bf 521} (2001) 133
  [arXiv:astro-ph/0105479].


\bibitem{PHSS}
  V.~Paschalidis, S.~M.~H.~Halataei, S.~L.~Shapiro and I.~Sawicki,
  Class.\ Quant.\ Grav.\  {\bf 28} (2011) 085006
  [arXiv:1103.0984 [gr-qc]].




\bibitem{WGA}
  B.~Wang, Y.~g.~Gong and E.~Abdalla,
  Phys.\ Lett.\  B {\bf 624} (2005) 141
  [arXiv:hep-th/0506069].


\bibitem{KLM}
  H.~Kim, H.~W.~Lee and Y.~S.~Myung,
  Phys.\ Lett.\  B {\bf 632} (2006) 605
  [arXiv:gr-qc/0509040].

 \bibitem{KLMb}
  K.~Y.~Kim, H.~W.~Lee and Y.~S.~Myung,
  Mod.\ Phys.\ Lett.\  A {\bf 22} (2007) 2631
  [arXiv:0706.2444 [gr-qc]].

  \bibitem{BGLe}
K.~Bamba, C.~Q.~Geng and C.~C.~Lee,
  JCAP {\bf 1008} (2010) 021
  [arXiv:1005.4574 [astro-ph.CO]].

\bibitem{Frolov}
  A.~V.~Frolov,
  Phys.\ Rev.\ Lett.\  {\bf 101} (2008) 061103
  [arXiv:0803.2500 [astro-ph]].










\end{thebibliography}

\end{document}